\documentclass[12pt,a4paper]{article}
\usepackage[utf8]{inputenc}
\usepackage{amsmath, amssymb, amsfonts}
\usepackage{graphicx}
\usepackage{hyperref}
\usepackage{cite}
\usepackage{geometry}
\usepackage{authblk}
\usepackage{orcidlink}
\title{Non-Markovian Quantum Decay in Complex Environments: A Hyperstatistical Approach}
\author{Nicola Fabiano\, \orcidlink{0000-0003-1645-2071}}
\affil{``Vin\v{c}a'' Institute of Nuclear Sciences - National 
Institute of the Republic of Serbia, University of Belgrade, Mike Petrovi\'{c}a 
Alasa 12--14, 11351 Belgrade, Serbia; nicola.fabiano@gmail.com}
\date{}

\begin{document}
\maketitle

\begin{abstract}
Standard Markovian frameworks, such as Fermi's Golden Rule, typically model the decay of an unstable quantum state as a simple exponential process, presuming a featureless environment. Nevertheless, when the system is immersed in a complex setting---marked by structural disorder, extended interactions, or intense fluctuations---local decay rates vary significantly. This variation gives rise to non-Markovian behavior and algebraic ``long-time tails.'' Here, we utilize the newly developed \textit{hyperstatistics} formalism to investigate quantum decay within these intricate surroundings. By assuming that local decay rates across mesoscopic regions follow a $\gamma$-distribution, we demonstrate that the overall survival probability takes the form of a $q$-exponential function. Furthermore, we employ the $q$-deformed Gamma function, constructed through the Mellin transform of the $q$-exponential, to evaluate the moments of the decay-time distribution. Our analysis reveals that the average quantum lifetime is given by $\langle t\rangle = \int_0^\infty P(t)\,dt= [\langle\Gamma\rangle(2-q)]^{-1}$, which remains finite provided $q<2$. Conversely, the finiteness of the second moment---and consequently the lifetime variance---demands the more restrictive bound $q<3/2$. These findings offer a nuanced physical picture: in the interval $1<q<3/2$, both the average lifetime and its variance are well-defined; for $3/2\le q<2$, the average lifetime is finite yet lifetime fluctuations exhibit infinite variance; finally, when $q\ge2$, the average lifetime itself diverges, signifying a genuinely localized, trapped, or Griffiths-type phase.
\end{abstract}

\section{Introduction}
Within conventional quantum mechanics, an unstable state coupled to a flat, memoryless continuum exhibits an exponential survival probability,
\begin{equation}
P(t)=e^{-\Gamma_0 t},
\end{equation}
where $\Gamma_0$ denotes the steady decay rate derived, for instance, via Fermi's Golden Rule. This standard picture hinges on the premise that the environment rapidly loses any memory of its interaction with the decaying system.

In contrast, embedding a quantum system within a \textit{complex environment}---such as a disordered semiconductor, a fractal phonon bath, or a medium exhibiting long-range correlations and memory---causes the local density of states and the system-environment coupling to fluctuate across space and time. Under these conditions, the environment generates a wide spectrum of decay rates, resulting in non-exponential, algebraic ``long-time tails'' in the survival probability. Conventional master equations frequently fail to encapsulate these prolonged memory effects without the inclusion of complicated non-Markovian kernels.

To address complex systems where Boltzmann-Gibbs statistics fails within localized mesoscopic regions, the \textit{hyperstatistics} paradigm was recently formulated \cite{Squillante2026Hyperstatistics, Squillante2026Remarks}. This approach incorporates a distribution of statistical weights across confined domains, naturally giving rise to $q$-exponential functions. Mathematically, this methodology shares deep connections with, and is often equivalent to, the \textit{superstatistics} framework developed by Beck and Cohen \cite{BeckCohen2003}. In superstatistics, a superposition of local Boltzmann-Gibbs distributions over a fluctuating intensive parameter yields $q$-exponential marginal distributions. For the quantum decay scenario explored here, the fluctuating parameter is the local decay rate $\Gamma$. In this work, we adopt the terminology and the $q$-Gamma-function machinery established in the hyperstatistics literature \cite{Squillante2026Hyperstatistics, Squillante2026Remarks}.

Although initially deployed to describe classical relaxation and high-energy phenomena \cite{Squillante2026Hyperstatistics}, the mathematical architecture of hyperstatistics is highly applicable to quantum dynamics. Notably, the original authors explicitly highlighted ``quantum decays'' and ``photoluminescence decay'' as primary domains for their framework \cite{Squillante2026Hyperstatistics, Squillante2026Remarks}.

The present article elaborates on the implementation of hyperstatistics to non-Markovian quantum decay. We translate fluctuating decay rates into the hyperstatistical language, deduce the $q$-exponential survival probability, and apply the $q$-generalized Gamma function to pinpoint the exact conditions required for the quantum state to maintain physically sensible, finite moments in its lifetime distribution.

\section{Quantum Decay in Fluctuating Environments}
Consider an unstable quantum state $|\psi\rangle$ situated within a complex environment. We divide the environment into discrete, localized mesoscopic \textit{domains} \cite{Squillante2026Hyperstatistics, Squillante2026Remarks}. Inside any given domain $i$, the environment is treated as locally Markovian, meaning the quantum survival probability undergoes standard exponential decay:
\begin{equation}
P_i(t)=e^{-\Gamma_i t},
\label{localdecay}
\end{equation}
where $\Gamma_i$ represents the decay rate specific to that domain.

In a complex setting, these local decay rates $\Gamma_i$ are not homogeneous. Driven by intrinsic disorder, multiplicative noise, or long-range correlations, $\Gamma_i$ varies from one mesoscopic domain to another. To model this heterogeneity, we assign a probability distribution function (PDF) to the local decay rates. Adhering to the core postulate of hyperstatistics \cite{Squillante2026Hyperstatistics}, we model the decay rates using a $\gamma$-distribution:
\begin{equation}
f(\Gamma)
=
\frac{1}{\Gamma(n)}
\left(\frac{n}{\langle\Gamma\rangle}\right)^n
\Gamma^{n-1}
\exp\!\left(-\frac{n\Gamma}{\langle\Gamma\rangle}\right),
\label{gammadist}
\end{equation}
where $\langle\Gamma\rangle$ denotes the macroscopic average decay rate, $n>0$ acts as a shape parameter reflecting the level of environmental complexity or disorder, and $\Gamma(n)$ is the standard Euler Gamma function.

The first two moments of the distribution \eqref{gammadist} are
\begin{equation}
\langle\Gamma\rangle_{\rm dist}=\langle\Gamma\rangle,
\qquad
{\rm Var}(\Gamma)=\frac{\langle\Gamma\rangle^2}{n}.
\end{equation}
Consequently, once we apply the identification $n=1/(q-1)$ detailed below, the parameter $q-1$ directly quantifies the relative variance of the local decay rates.

\section{Hyperstatistical Formulation of the Survival Probability}
The observable, macroscopic survival probability $P(t)$ is obtained by statistically averaging over all mesoscopic domains. Mathematically, this corresponds to taking the Laplace transform of the local decay probabilities, weighted by the $\gamma$-distribution:
\begin{equation}
P(t)
=
\int_0^{\infty} e^{-\Gamma t} f(\Gamma)\,d\Gamma
=
\frac{1}{\Gamma(n)}
\left(\frac{n}{\langle\Gamma\rangle}\right)^n
\int_0^{\infty}
\Gamma^{n-1}
\exp\!\left[-\Gamma\left(t+\frac{n}{\langle\Gamma\rangle}\right)
\right]d\Gamma.
\end{equation}
Solving this integral results in a closed-form algebraic expression:
\begin{equation}
P(t)
=
\left[
1+\frac{\langle\Gamma\rangle t}{n}
\right]^{-n}.
\end{equation}
To link this outcome with the framework of nonadditive $q$-statistical mechanics \cite{Tsallis1988}, we define the entropic index $q>1$ via the relationship
\begin{equation}
n=\frac{1}{q-1}.
\end{equation}
Substituting this into our expression, the macroscopic survival probability elegantly converts into a \textit{$q$-exponential function} \cite{Squillante2026Hyperstatistics}:
\begin{equation}
P(t)
=
\left[
1+(q-1)\langle\Gamma\rangle t
\right]^{-\frac{1}{q-1}}
\equiv
\exp_q(-\langle\Gamma\rangle t).
\label{qsurv}
\end{equation}
This mathematical structure mirrors the $q$-exponential relaxation laws previously established in superstatistics and hyperstatistics studies \cite{Squillante2026Hyperstatistics, BeckCohen2003}.

The physical consequences of Eq.~\eqref{qsurv} can be summarized as follows:
\begin{itemize}
\item \textbf{Markovian limit ($q\to 1$):} When the environment's complexity disappears, $n\to\infty$ and $q\to1$. The $q$-exponential seamlessly reverts to the classic Markovian exponential decay:
\begin{equation}
P(t)\to e^{-\langle\Gamma\rangle t}.
\end{equation}
\item \textbf{Non-Markovian long-time tails ($q>1$):} In complex environments, $q>1$. As $t\to\infty$, the survival probability displays a power-law tail:
\begin{equation}
P(t)\sim t^{-\frac{1}{q-1}}.
\end{equation}
This algebraic decay signifies effective quantum memory and non-Markovian behavior induced by the heterogeneous surroundings. Thus, the hyperstatistical approach offers a rigorous, analytically manageable substitute for embedding arbitrary integro-differential non-Markovian kernels.
\end{itemize}

\section{Mean Lifetime, Second Moment, and the $q$-Generalized Gamma Function}
A fundamental issue in the study of quantum decay is determining whether the unstable state maintains a well-defined, finite average lifetime. Given a survival probability $P(t)$, the complementary cumulative probability is $1-P(t)$, and the decay-time probability density is its negative derivative:
\begin{equation}
p(t)=-\frac{dP(t)}{dt}.
\end{equation}
The average lifetime is therefore expressed as
\begin{equation}
\langle t\rangle
=
\int_0^\infty t\,p(t)\,dt
=
\int_0^\infty t\left(-\frac{dP}{dt}\right)dt.
\end{equation}
Through integration by parts, and assuming the boundary term $tP(t)\to0$ as $t\to\infty$, we arrive at the standard survival-probability integral:
\begin{equation}
\langle t\rangle
=
\int_0^\infty P(t)\,dt.
\label{meanlifetime}
\end{equation}
This distinction is crucial. The integral $\int_0^\infty tP(t)\,dt$ does not represent the mean lifetime; rather, it carries dimensions of time squared and relates to the second moment of the decay-time distribution. Specifically, provided the second moment exists, we have:
\begin{equation}
\langle t^2\rangle
=
\int_0^\infty t^2 p(t)\,dt
=
2\int_0^\infty tP(t)\,dt.
\label{secondmomentrelation}
\end{equation}
Applying the survival probability from Eq.~\eqref{qsurv}, the average lifetime becomes
\begin{equation}
\langle t\rangle
=
\int_0^{\infty}
\exp_q(-\langle\Gamma\rangle t)\,dt.
\end{equation}
By introducing the substitution $x=\langle\Gamma\rangle t$ (so $dt=dx/\langle\Gamma\rangle$), the integral transforms to
\begin{equation}
\langle t\rangle
=
\frac{1}{\langle\Gamma\rangle}
\int_0^{\infty}
\exp_q(-x)\,dx.
\label{meanlifex}
\end{equation}
At this stage, we introduce the \textit{$q$-generalized Gamma function}, $\Gamma(n_{\rm mellin},q)$, which is defined as the Mellin transform of the $q$-exponential \cite{Squillante2026Hyperstatistics}:
\begin{equation}
\Gamma(n_{\rm mellin},q)
=
\int_0^{\infty}
x^{n_{\rm mellin}-1}
\exp_q(-x)\,dx.
\label{qgammadef}
\end{equation}
For $q>1$, this integral admits an exact analytical evaluation:
\begin{equation}
\Gamma(n_{\rm mellin},q)
=
\frac{1}{(q-1)^{n_{\rm mellin}}}
\frac{
\Gamma(n_{\rm mellin})
\Gamma\!\left(
\frac{1}{q-1}-n_{\rm mellin}
\right)
}{
\Gamma\!\left(
\frac{1}{q-1}
\right)
}.
\label{gmellin}
\end{equation}
The requirement for convergence stems from the second Gamma function in the numerator:
\begin{equation}
\frac{1}{q-1}-n_{\rm mellin}>0
\quad\Longrightarrow\quad
q<1+\frac{1}{n_{\rm mellin}}.
\label{generalconvergence}
\end{equation}
For the average lifetime integral \eqref{meanlifex}, the exponent of $x$ is zero, implying
\begin{equation}
n_{\rm mellin}-1=0
\quad\Longrightarrow\quad
n_{\rm mellin}=1.
\end{equation}
Thus, we can write
\begin{equation}
\langle t\rangle
=
\frac{\Gamma(1,q)}{\langle\Gamma\rangle}.
\end{equation}
Evaluating Eq.~\eqref{gmellin} at $n_{\rm mellin}=1$ yields
\begin{equation}
\Gamma(1,q) =
\frac{1}{q-1}
\frac{\Gamma(1)\Gamma\!\left(\frac{1}{q-1}-1\right)
}{
\Gamma\!\left(
\frac{1}{q-1}
\right)
}
=
\frac{1}{q-1}
\frac{1}{
\frac{1}{q-1}-1
}
=
\frac{1}{2-q}.
\end{equation}
Consequently, the average lifetime is
\begin{equation}
\langle t\rangle
=
\frac{1}{\langle\Gamma\rangle(2-q)}.
\label{meanlifetimefinal}
\end{equation}
This result possesses the correct physical dimensions of time. In the Markovian limit ($q\to1$), it simplifies to
\begin{equation}
\langle t\rangle = \frac{1}{\langle\Gamma\rangle},
\end{equation}
which is the expected outcome for pure exponential decay.

Based on Eq.~\eqref{generalconvergence} with $n_{\rm mellin}=1$, the convergence criterion for the average lifetime is
\begin{equation}
q<2.
\label{meanconvergence}
\end{equation}
Therefore, the average lifetime remains finite for $1<q<2$ and diverges when $q\ge2$.

Evaluating the second moment is also highly instructive, as it illuminates the unique significance of $q=3/2$. The integral $\int_0^\infty tP(t)\,dt$ maps to $n_{\rm mellin}=2$:
\begin{equation}
\int_0^\infty tP(t)\,dt
=
\frac{1}{\langle\Gamma\rangle^2}
\int_0^\infty x\exp_q(-x)\,dx
=
\frac{\Gamma(2,q)}{\langle\Gamma\rangle^2}.
\end{equation}
Utilizing Eq.~\eqref{secondmomentrelation}, the second moment of the lifetime distribution is
\begin{equation}
\langle t^2\rangle
=
\frac{2\Gamma(2,q)}{\langle\Gamma\rangle^2}.
\end{equation}
Setting $n_{\rm mellin}=2$ in Eq.~\eqref{gmellin} gives
\begin{equation}
\Gamma(2,q)
=
\frac{1}{(q-1)^2}
\frac{
\Gamma(2)
\Gamma\!\left(
\frac{1}{q-1}-2
\right)
}{
\Gamma\!\left(
\frac{1}{q-1}
\right)
}
=
\frac{1}{(2-q)(3-2q)}
=
\frac{1}{6-7q+2q^2}.
\end{equation}
Hence, we find
\begin{equation}
\langle t^2\rangle
=
\frac{2}{
\langle\Gamma\rangle^2(6-7q+2q^2)
}.
\label{secondmomentfinal}
\end{equation}
From Eq.~\eqref{generalconvergence} with $n_{\rm mellin}=2$, the convergence bound for the second moment is
\begin{equation}
q<\frac{3}{2}.
\label{varianceconvergence}
\end{equation}
The lifetime variance, defined as $\sigma_t^2 = \langle t^2\rangle-\langle t\rangle^2$, can be evaluated using Eqs.~\eqref{meanlifetimefinal} and \eqref{secondmomentfinal}:
\begin{equation}
\sigma_t^2
=
\begin{cases}
\displaystyle
\frac{1}{
\langle\Gamma\rangle^2(2-q)^2(3-2q)
},
& 1<q<\dfrac{3}{2},
\\[4mm]
+\infty,
& \dfrac{3}{2}\le q<2,
\\[2mm]
\text{not defined, since } \langle t\rangle \text{ diverges},
& q\ge2.
\end{cases}
\label{variancefinal}
\end{equation}
It is important to emphasize that $q=3/2$ does not mark the onset of an infinite average lifetime. Instead, it denotes the threshold where the lifetime variance, or equivalently the second moment of the decay-time distribution, diverges. The actual boundary for the divergence of the average lifetime is $q=2$.

\subsection{Convergence Criteria and Physical Regimes}
The derived convergence conditions delineate three unique physical regimes for a quantum state immersed in a complex environment:
\begin{enumerate}
\item \textbf{Finite lifetime and finite fluctuations ($1<q<3/2$):}
Both the average lifetime $\langle t\rangle$ and the variance $\sigma_t^2$ are bounded. The system displays non-Markovian power-law tails but ultimately decays, and the lifetime distribution possesses a well-defined spread.
\item \textbf{Finite mean lifetime but divergent variance ($3/2\le q<2$):}
The average lifetime $\langle t\rangle=[\langle\Gamma\rangle(2-q)]^{-1}$ stays finite, yet the second moment and variance blow up. Physically, this signifies a phase of extraordinarily broad lifetime fluctuations: while the average decay time is well-defined, the sample-to-sample or domain-to-domain variations are so heavily tailed that the variance becomes infinite. This is not a completely frozen state, but rather a heavily tailed, strongly disordered dynamical phase.
\item \textbf{Infinite mean lifetime / truly trapped regime ($q\ge2$):}
The integral for the average lifetime diverges. Physically, this describes a state of extreme environmental complexity, such as profound Anderson localization, deep trapping, or a Griffiths-like phase \cite{Squillante2026Hyperstatistics, Griffiths1969}. In this regime, the quantum state is practically frozen over experimental timescales, and the survival probability decays too sluggishly to be integrable.
\end{enumerate}

\section{Concrete Application: Photoluminescence Decay in Quantum Dot Ensembles}
To anchor this theoretical construct in a practical experimental context, we examine the photoluminescence (PL) decay dynamics of quantum dot (QD) ensembles situated in disordered solid matrices or in proximity to interfaces. For an ideal, isolated QD, the radiative recombination of an exciton exhibits a straightforward exponential decay. In real-world settings, however, the local density of optical states (LDOS), phonon bath coupling, surface defects, and spectral diffusion undergo significant fluctuations from one dot to another, or even temporally within a single dot. These variations generate a mesoscopic landscape where the decay rate $\Gamma$ is not a solitary constant but rather follows a statistical distribution.

By applying the hyperstatistics framework, the macroscopic PL decay signal---or the excited-state population driving the measured PL transient---is obtained by averaging over these domains. When the recorded transient is proportional to the excited-state population, or following standard normalization of the decay curve, one naturally derives
\begin{equation}
I(t)=I_0\exp_q(-\langle\Gamma\rangle t).
\end{equation}
This formulation directly accounts for the pervasive non-exponential, power-law tails observed in QD luminescence, eliminating the need for purely empirical stretched-exponential Kohlrausch fits. Recent statistical investigations into PL decay kinetics in QD ensembles have explicitly shown that this hyperstatistical modeling effectively captures the influence of inorganic shell composition and environmental disorder, confirming the $q$-exponential decay as a fundamental marker of complex quantum relaxation \cite{Martins2022, Squillante2026Remarks,Breuer2002}.

The moment analysis holds particular value for experimentalists. The mean PL lifetime should be determined from the area under the survival or population decay curve, $\langle t\rangle=\int_0^\infty P(t)\,dt,$ while the breadth or fluctuation of the lifetime distribution is dictated by the second moment. In highly disordered QD ensembles, researchers might therefore detect a finite average lifetime accompanied by exceptionally broad, potentially effectively divergent, lifetime fluctuations when $3/2\le q<2$.

\section{Conclusions}
We have shown that the \textit{hyperstatistics} paradigm delivers an exact, analytically manageable resolution to the challenge of non-Markovian quantum decay within complex environments. By representing the environment as an ensemble of mesoscopic domains featuring a $\gamma$-distribution of local decay rates, the macroscopic survival probability organically emerges as a $q$-exponential function.

Importantly, by leveraging the $q$-generalized Gamma function $\Gamma(n_{\rm mellin},q)$, we have pinpointed the exact convergence criteria for the moments of the lifetime distribution. The average quantum lifetime is $\langle t\rangle=[\langle\Gamma\rangle(2-q)]^{-1}$, remaining finite for $q<2$. The second moment, and thus the lifetime variance, converges exclusively for $q<3/2$. As a result, the interval $3/2\le q<2$ must be understood as a heavy-tailed fluctuation phase featuring a finite average lifetime but infinite variance, whereas the domain $q\ge2$ signifies a genuinely trapped or localized state with an infinite average lifetime. As illustrated by practical applications like the photoluminescence decay in quantum dot ensembles, hyperstatistics stands out as a robust and adaptable instrument for quantum statistical mechanics, seamlessly connecting microscopic decay processes with macroscopic complex behaviors.

\end{document}